\documentclass[10pt,aps,prl,twocolumn,floatfix]{revtex4}

\usepackage{graphicx}
\begin{document}
\baselineskip 11pt

\title{\large \bf Generating Generalized Distributions from Dynamical Simulation}

\author{Eric J. Barth}
\email{barth@kzoo.edu}
\affiliation{Department of Mathematics\\ Kalamazoo College, Kalamazoo, MI 49006, USA}
\author{Brian B. Laird} 
\email{blaird@ku.edu}
\affiliation{Department of Chemistry\\ University of Kansas, Lawrence, KS 66045, USA}
\author{Benedict J. Leimkuhler}
\email{bl12@mcs.le.ac.uk}
\affiliation{Department of Mathematics and Computer Science\\University of Leicester,
Leicester LE1 7RH, UK}

\begin{abstract}
We present a general molecular-dynamics simulation scheme, based
on the Nos\'e thermostat, for sampling according to arbitrary  
phase space distributions.  We formulate numerical methods based on both
Nos\'{e}-Hoover and Nos\'{e}-Poincar\'{e} thermostats 
for two specific classes of distributions; namely, those that
are functions of the system Hamiltonian and those for which position
and momentum are statistically independent. 
As an example, we propose a generalized variable temperature distribution that
is designed to accelerate sampling in molecular systems.

\end{abstract}

\maketitle
\section{Introduction}

Molecular-dynamics (MD) computer simulation is a widely used tool
in biology, chemistry, physics and materials science\cite{Frenkel02,Allen87}. 
Much of the power in the technique lies in the ability to generate phase-space
trajectories weighted according to a relevant statistical-mechanical distribution.
In the first MD simulations, straightforward integration of the equations of motion
for the system under study yielded energy conserving trajectories that, 
assuming ergodicity, generated 
microcanonical (constant $NVE$) equilibrium distributions of phase-space configurations. 
Later, to better mimic experimental conditions, a variety of MD techniques were developed that 
generate phase-space distributions according to other standard statistical-mechanical 
distributions, such as canonical ($NVT$)\cite{Nose84a,Hoover85,Bond99}, isothermal-isobaric 
($NPT$)\cite{Andersen80,Sturgeon00}, and grand-canonical$(\mu VT)$\cite{Lynch97}. Recently, 
however, there has been growing interest in the simulation of
systems with distributions that go beyond textbook statistical mechanical
ensembles. For example, molecular-dynamics methods for the simulation of systems obeying
Tsallis statistics\cite{Tsallis88} have been developed by  Plastino and 
Anteneodo\cite{Plastino97} and Andriciaoaei and Straub\cite{Andricioaei97}. In this work 
we outline a {\it general} molecular-dynamics scheme, based on the Nos\'e 
thermostat\cite{Nose84a,Nose84b}, to generate configurations according to an
arbitrary 
phase-space distribution. 

A primary motivation for the development of algorithms for the generation 
of non-standard distributions is the need for methods that accelerate the configurational
sampling of systems. Many systems are not sufficiently ergodic on the time scale of standard 
molecular-dynamics simulations to ensure the convergence of statistical averages.  
This is especially true of macromolecules, biomolecules and amorphous materials. 
Over the past decade, a number of methods have been
developed to enhance sampling in MD.  Berne and Straub\cite{Berne97} have recently written 
an excellent review of new sampling methods.  Central to these approaches has been the 
recognition that high activation barriers cause a bottleneck in phase space, 
rendering transitions between states unlikely.   A common thread among many
methods is the systematic deformation of the potential (or total) energy surface 
to accelerate barrier crossing, either by lowering the barriers or raising the potential 
valleys.  From a statistical mechanical perspective, such energy modifications 
induce a corresponding modification in the phase-space distribution by enhancing the statistical
weight of configurations in the vicinity of energy barriers. 
Explicit knowledge of the modified sampling distribution allows for
statistical reweighting of the computed trajectories
to achieve averages in the original ensemble.

The simplest method for enhancing sampling is to scale the full Hamiltonian
by some factor less than unity. This is equivalent to performing the
simulation at a higher  temperature. If averages are desired at temperature $T$, 
isothermal MD simulations can be carried out at some higher temperature $T^*$, 
with averages at the original temperature
computed by reweighting the probability of each configuration
by a factor of $\exp{\left [ \left (\frac{1}{kT} - \frac{1}{kT^*}\right ) H(p,q)\right ]}$. 
A significant disadvantage with such temperature boost approaches 
is that, unless the boost is sufficiently small, low energy configurations are
not visited with a frequency large enough to obtain acceptable statistics.
A related approach, Multicanonical MD, is based on a Monte Carlo technique 
of the same name\cite{Berg91} and 
uses preliminary high temperature trajectories to construct 
a distribution that is nearly flat in the position coordinates, allowing
nearly uniform sampling of coordinate space in subsequent simulations.  
Multicanonical MD has been demonstrated to accelerate conformational
sampling in model polypeptides\cite{Nakajima97} and atomic clusters.
 
Another approach is
Voter's hyperdynamics\cite{Voter96,Voter97}, which employs a ``boost potential''
to reduce the sampling probability in low energy regions, 
thereby accelerating barrier crossing 
due to diminished relative energetic cost.  With boost potentials
chosen to leave the potential energy in
barrier regions unchanged, transition state theory
arguments can be used to obtain good transition rate statistics.
However, the low-energy sampling problem remains as in high temperature dynamics.
Hyperdynamics has been used successfully in solid state systems,
but the method is not generally applicable to liquids,
where the presence of many saddle points hampers
the identification of well-defined barrier regions.

Among recent methods of enhanced sampling, 
the approach of Straub and co-workers\cite{Andricioaei97}
is of particular interest,
with Monte Carlo and MD methods 
based on potential energy modifications
that sample coordinates from alternative densities  
according to a formalism
motivated by the non-extensive Tsallis entropy\cite{Tsallis88}.
The Tsallis-Straub approach is easy to implement, amounting
to a simple modification of the interaction forces according
to the gradient of an effective potential.
Recently, a more direct application of Tsallis entropy to
MD was suggested by Plastino and Anteneodo\cite{Plastino97}. Based on the idea of
an effective Hamiltonian, these authors showed that canonical
sampling with respect to the effective Hamiltonian is
equivalent to Tsallis sampling.  Significantly, this work considered
only the Tsallis regime in which coordinate sampling is restricted
to low energy regions.

In this paper we present a dynamical framework for sampling according to a general 
class of probability density functions, including but not limited to the Tsallis density.
In order to introduce the idea of sampling from non-microcanonical distributions and to
provide the necessary background for our generalized dynamics we discuss in 
Section~\ref{nose} the extended Hamiltonian approach of Nos\'{e}\cite{Nose84a} to canonical
(constant temperature) sampling, as well as the Nos\'{e}-Hoover\cite{Hoover85} 
and Nos\'e-Poincar\'e\cite{Bond99}
approaches for implementing real-time formulations of Nos\'e dynamics.
In Section~\ref{gdd} we introduce Generalized 
Distribution Dynamics (GDD) and discuss the technique 
for two special classes of systems: those for which the position and momentum 
distributions are separable and those for which the phase space distribution is a function
of the full Hamiltonian, and  
show how the Nos\'{e}
framework can be used to derive
the equations of motion that produce trajectories that sample from generalized distributions. 
In section \ref{tpd} we present as an example the variable
temperature distribution for accelerating the sampling of systems with
high barriers. Numerical experiments on a double well potential using both the separable 
and full Hamiltonian GDD approaches are presented in section~\ref{numexp}. 

\section{Sampling from a Canonical Distribution: the Nos\'e Thermostat} \label{nose}

In traditional (NVE) MD simulation, the equations of motion corresponding to the
system Hamiltonian, $H({\bf p},{\bf q})$, are integrated to generate the trajectories. 
The trajectory is constrained to the constant energy surface, 
$E = H({\bf p},{\bf q})$ determined by the initial values of coordinates and
momenta.  States in phase space along solutions are said to be sampled from the
microcanonical, or constant energy, distribution according to the
probability density $\rho_{\rm NVE}({\bf q},{\bf p})$ that is proportional to
$\delta(H({\bf q},{\bf p})-E),$ where $\delta$ is the Dirac delta function.


Due in part to a desire to
bring simulation into accord with laboratory experiments that are
typically conducted at some fixed temperature, methods have been developed for
generating trajectories which sample from the canonical, or
constant temperature, ensemble according to the probability density 
$\rho_{\rm NVT}({\bf q},{\bf p})$,
which is proportional to $\exp{[-\beta H({\bf q},{\bf p})]}$
where $\beta = 1/(k_BT)$, $T$ being the temperature and $k_B$ the
Boltzmann constant.  In contrast to the microcanonical case, canonical sampling
allows states at all energies, though higher energy states have
lower probabilities depending on the value of temperature $T$.

 
Although other methods exist, the most widely used techniques for generating 
canonically distributed trajectories in MD simulation are based on the extended
Hamiltonian of Nos\'e\cite{Nose84a,Nose84b}: 
\begin{equation}
H_{Nos \acute{e}} = \frac{{\tilde{\bf p}}^T{\bf M}^{-1}{\tilde{\bf p}}}{2s^2} + V({\bf
  q})+ \frac{\pi^2}{2Q}+gk_BT \ln s, 
\label{noseHam}
\end{equation}
where $s$ and $\pi$ are conjugate thermostat variables, $Q$ is a fictional
thermostat mass which determines the strength of thermal coupling to the system,
$g = N_f+1$ (with $N_f$ being the number of degrees of freedom in the
system) and $\tilde{\bf p}$ is a virtual momentum related to
the actual momentum of the system by $\tilde{\bf p} = s {\bf p}$\cite{Nose84a}. 
The equations of motion generated by the Nos\'e Hamiltonian (Eq.~\ref{noseHam}) are
\begin{eqnarray}
\frac{d {\bf q}}{d \tau} &=& {\bf M}^{-1}{\tilde{\bf p}}/s^2 \label{n1} \\
\frac{d \tilde{\bf p}}{d \tau} &=&   -\nabla V({\bf q}) \label{n2}\\
\frac{d s}{d \tau} &=& \frac{\pi}{Q}  \label{n3}\\
\frac{d \pi}{d \tau} &=& \frac{\tilde{\bf p}^T{\bf M}^{-1}{\tilde{\bf p}}}{s^3}-gk_BT/s.  \label{n4}
\end{eqnarray}

The Nos\'e method regulates the temperature of the system through a dynamical time
transformation given by $\frac{d \tau}{d t} = s$, where $\tau$ is the
Nos\'e (virtual) time and $t$ is real time. 
The remarkable property of  Nos\'{e} dynamics is that microcanonical
sampling of the extended phase space $\{{\bf q},\tilde{\bf p},s,\pi\}$
yields canonical sampling in the reduced phase space,$\{{\bf q},{\bf p}\}$, provided
that the system is ergodic.  

For practical calculations of averages
such as velocity autocorrelation functions, it is convenient to work in a  
real time implementation of the Nos\'e 
thermostat.
The most commonly used real-time modification is due to
Hoover\cite{Hoover85}. Hoover recognized that one can generate a set of real-time
equations of motion by making the following transformations
to the Nos\'e equations of motion
\begin{enumerate}
\item change of variables: ${\bf p}=\tilde{\bf p}/s$.
\item time transformation: $d\tau / dt = s$,
\item change of variables:  $\eta=\ln s$ and $\xi = \dot{\eta}$. 
\end{enumerate}
The result is the following time-reversible system of equations, 
known as the Nos\'{e}-Hoover (NH) equations\cite{Hoover85}:
\begin{eqnarray}
\dot{\bf q} &=& {\bf M}^{-1}{\bf p} \label{nh1} \\
\dot{\bf p} &=&   -\nabla V({\bf q}) - \xi{\bf p} \label{nh2}\\
\dot{\eta} &=& \xi  \label{nh3}  \\
\dot{\xi} &=& \frac{1}{Q}\left [{\bf p}^T{\bf M}^{-1}{\bf p}-gk_BT\right ],  \label{nh4}
\end{eqnarray}
where $g=N_f$, the number of degrees of freedom in the system.
These equations of motion are non-Hamiltonian in form since the coordinate transformations
were not canonical; however, a conserved energy does exist given by
\begin{equation}
E_{NH} = \frac{{\bf p}^T{\bf M}^{-1}{\bf p}}{2} + V({\bf q})+ \frac{Q\xi^2}{2}+gk_BT 
\eta \;. \label{E0}
\end{equation}
(Although the variable $\eta$ has been 
decoupled from the system, it is helpful to include it in the
calculations so that $E$ can be monitored as an indicator
of trajectory stability.) 

Recently, Bond, Leimkuhler and Laird\cite{Bond99} have developed an
alternative real-time Nos\'e thermostat scheme, the Nos\'{e}-Poincar\'{e} method,
which is Hamiltonian in form, allowing for the use of symplectic integration schemes 
(which have been shown to give superior stability in long time simulation\cite{Sanz-Serna95}).  
This is accomplished by performing a time transformation, not to the Nos\'e equations of motion as
with Nos\'e-Hoover, but directly to the Hamiltonian using a Poincar\'{e} time 
transformation, as follows:
\begin{equation}
H_{NP} = s(H_{Nos \acute{e}} - H_0),
\label{Trans}
\end {equation}
where $H_0$ is the initial value of $H_{Nos \acute{e}}$.
It can be easily verified\cite{Bond99} that the phase space trajectory generated by $H_{NP}$ is identical 
to that generated by $H_{Nos \acute{e}}$ except for a time transformation $\frac{d \tau}{d t} = s$.
The Nos\'e-Poincar\'e equations of motion are 
\begin{eqnarray}
\dot{\bf q} &=& s^{-1} {\bf M}^{-1}{\tilde{\bf p}} \\
\label{npeom1}
\dot{s}& =& \frac{s \pi}{Q} \\
\label{npeom2}
\dot{\tilde{\bf p}} &= &-s \nabla V(q) \\
\label{npeom3}
\dot{\pi} &=& s^{-2} \tilde{\bf p}^T {\bf M}^{-1} \tilde{\bf p} 
- gkT -\Delta{H} \;,
\label{npeom4}
\end{eqnarray}
where
\begin{equation}
\Delta{H} = \frac{\tilde{\bf p}^T \tilde{\bf M}^{-1} 
\tilde{\bf p}}{2 s^2} + V_c(q) + \frac{\pi^2}{2Q} + gkT\ln s - {H}_0 \;. 
\label{npeom5}
\end{equation}
Note that, the exact solution to Nos\'e-Poincar\'e equations of motion generates trajectories that
are identical to those generated by the Nos\'e-Hoover scheme, exactly solved.  It is in 
the construction of approximate numerical methods that these two approaches differ.

Although we favor the Nos\'{e}-Poincar\'{e} method in all cases,  
the Nos\'e-Hoover formalism is more familiar within the simulation community.   
For this reason, in the current article, we present schemes based on
both the Nos\'{e}-Hoover and Nos\'{e}-Poincar\'{e} approaches.  

In certain systems, for example those with few particles or strong
harmonic components, the ergodicity assumption basic to the Nos\'e
approaches is not met.  For these cases, the notion of Nos\'{e}-Hoover
chains has been developed\cite{Martyna92}, in which the Hamiltonian is further
extended with additional thermostat variables that are coupled to each other.  It has
been demonstrated that NH chains, with properly chosen thermostat
masses, can induce the needed ergodicity so that NH dynamics
provides a means of sampling from the canonical distribution.
We discuss NH chains further in section \ref{numexp} and in the Appendix.

\section{Generalized Distribution Dynamics\label{gdd}}
In this section we present a
dynamical  scheme for sampling points in phase space according to
general function
$F({\bf p},{\bf q})$  
which satisfies the properties
of a probability density function in the phase space variables
$\{{\bf p},{\bf q}\}$:
$$ \int_{{\bf p},{\bf q}} F({\bf p},{\bf q}) = 1 \mbox{     and     } F({\bf p},{\bf
  q}) \ge 0.$$
In analogy to the procedure used by Plastino and Anteneodo\cite{Plastino97} to
develop an MD method to generate the canonical Tsallis distribution, 
we relate the general density to the canonical density by way
of an effective Hamiltonian $H_{\rm eff}$ as
$$F({\bf p},{\bf q}) = e^{-\beta H_{\rm eff} },$$
which yields
\begin{eqnarray}
H_{\rm eff}=-\frac{1}{\beta}\ln F({\bf p}, {\bf q}). \label{Heff}
\end{eqnarray}
It is clear that canonical sampling with respect to the effective Hamiltonian
is equivalent to sampling according to the generalized
probability density $F$.  To achieve canonical sampling with $H_{\rm eff}$ 
we write the Nos\'{e} Hamiltonian for Generalized Distribution Dynamics:
\begin{equation}
H^F_{Nos \acute{e}} = -\frac{1}{\beta}\ln F(\tilde{\bf p}/s, {\bf q})
+ \frac{\pi^2}{2Q}+gk_BT \ln
s. \label{gddnose}
\end{equation}

From the equations of motion generated from this Nos\'e Hamiltonian and after applying 
the transformations described in the previous section, we
obtain the Nos\'e-Hoover GDD equations of motion:
\begin{eqnarray}
\dot{\bf q} &=& -\frac{k_BT}{F({\bf p},{\bf q})}\nabla_{\bf p} \;
 F({\bf p}, {\bf q}) \label{gddnh1} \\
\dot{\bf p} &=&  \frac{k_BT}{F({\bf p}, {\bf q})}\nabla_{\bf q} \; F({\bf p}, {\bf q})
 -\xi{\bf p} \label{gddnh2} \\
\dot{\eta} &=& \xi  \label{gddnh3}\\
\dot{\xi} & =& \frac{1}{Q} \left [\frac{-k_BT}{F({\bf p}, {\bf q})}{\bf p}^T
\nabla_{\bf p} \; F({\bf p}, {\bf q}) - gk_BT\right ]. \label{gddnh4} 
\end{eqnarray}

Similarly, the Nos\'e-Poincar\'e equations of motion for GDD are
\begin{eqnarray}
\dot{\bf q} &=& -\frac{kT}{F(\tilde{\bf p}/s,{\bf q})} \nabla_{\tilde{\bf p}/s}
F(\tilde{\bf p}/s,{\bf q}) \\
\dot{\tilde{\bf p}} &=& \frac{kT}{F(\tilde{\bf p}/s,{\bf q})} \nabla_{\tilde{\bf q}}
F(\tilde{\bf p}/s,{\bf q}) \\
\dot{\bf s} &=& \frac{s \pi}{Q} \\
\dot{\bf \pi} &=& \frac{kT}{s F(\tilde{\bf p}/s,{\bf q})} \nabla_{\tilde{\bf p}/s}
F(\tilde{\bf p}/s,{\bf q}) - gkT \nonumber \\ 
&& - \Delta H_{Nos \acute{e}}  \;.
\end{eqnarray}

These formulations disrupt the separability of variables present in
the original NH and NP equations of motion [Eqs. (\ref{nh1})--(\ref{nh4}) and 
(\ref{npeom1}--\ref{npeom5}), respectively].  A time reversible
discretization of the GDD equations would involve the solution of nonlinear 
equations in ${\bf q}$ and ${\bf p}$ at every step.  Iterative solution would require
many evaluations of the potential energy and its gradient at each
step, likely adding tremendously to the computational burden.  
We address this issue by considering two special classes of probability density u
functions that maintain variable separability:

\vspace{0.2in}
\noindent
{\it Case 1: GDD for Separable Distribution Functions} 

Consider separable probability distribution functions of the form
$$F({\bf p},{\bf q}) =A({\bf p})B({\bf q}).$$
We can relate the separable density to the canonical density by way
of effective kinetic and potential energies $K_{\rm eff}$ and
$V_{\rm eff}$ as
$$F({\bf p},{\bf q}) = e^{-\beta K_{\rm eff} }e^{-\beta V_{\rm eff} },$$
leading to
\begin{equation}
K_{\rm eff}({\bf p})=-\frac{1}{\beta}\ln A({\bf p})  \;;\;\;
V_{\rm eff}({\bf q})=-\frac{1}{\beta}\ln B({\bf q}). \label{KVeff}
\end{equation}
Canonical sampling
with respect to the effective Hamiltonian
$H_{\rm eff}=K_{\rm eff}+V_{\rm eff}$ 
is equivalent to sampling according to the generalized
probability density $F$.  
Following the procedure outlined in the previous section, canonical sampling with 
$H_{\rm eff}$ can be achieved using the Nos\'e-Hoover GDD equations  of 

motion, which for a separable distribution function are obtained as:
\begin{eqnarray}
\dot{\bf q} &=& \nabla_{\bf p} K_{\rm{eff}}({\bf p}) \label{gddsnh1} \\
\dot{\bf p} &=&  -\nabla_{\bf q} V_{\rm{eff}}({\bf q}) -\xi {\bf p} \label{gddsnh2} \\
\dot{\eta} &=& \xi \label{gddsnh3}\\
\dot{\xi} &=& \frac{1}{Q} \left [{\bf p}^T \nabla_{\bf p} K_{\rm eff}({\bf p}) - gk_BT \right ] \;.
\label{gddsnh4}
\end{eqnarray}
Generation of the Nos\'e-Poincar\'e equations of motion for this class of distributions
follows similarly. 

Note that these equations have a simple relationship with 
the NH equations (\ref{nh1})--(\ref{nh4}).
Any existing implementation of the NH (or NP) equations of motion can be easily modified
for separable GDD by
the replacement of ${\bf M}^{-1}{\bf p}$ 
by $\nabla K_{\rm eff}({\bf p})$ in equations (\ref{nh1}), 
(\ref{nh4}), and $V({\bf q})$ by $V_{\rm eff}({\bf q})$ 
in equation (\ref{nh2}).  

The most important applications for GDD for separable distributions are those in which 
only the coordinate distribution is altered through modification of the potential. Such 
potential-only modifications are at the heart of Voter dynamics\cite{Voter97} and the Tsallis
statistics based methods for accelerated sampling of Straub and 
Andricioaei\cite{Andricioaei97}.  For such systems $K_{\rm eff}$ is equal to its standard
form $\frac{1}{2} {\bf p}^T {\bf M}^{-1} {\bf p}$ and $V_{\rm{eff}}$ is given by
eq.~(\ref{KVeff}).  Implementation of GDD for such systems is straightforward as any
existing Nos\'e-Hoover (or Nos\'e-Poincar\'e) code could be used without modification
(other than the use of a modified input potential surface). 

\vspace{0.2in}
\noindent
{\it Case 2: GDD for distributions that are functions of the Hamiltonian} 

Here we consider distributions that are
formal functions of the scalar Hamiltonian:  $F(H({\bf p},{\bf q}))$.
Defining the effective Hamiltonian as
$$H_{\rm eff}= (-1/\beta)\ln F(H({\bf p},{\bf q})) \equiv f(H({\bf p},{\bf q}))$$ with
associated Nos\'{e} Hamiltonian, 
the Nos\'e-Hoover GDD equations of motion for this case are
\begin{eqnarray}
\dot{\bf q} &=& f'(H({\bf p},{\bf q})){\bf M}^{-1}{\bf p} \label{gnh1} \\
\dot{\bf p} &=&   
-f'(H({\bf p},{\bf q}))\nabla V({\bf q}) - \xi \bf p \label{gnh2}\\
\dot{\eta} &=& \xi  \label{gnh3}\\
\dot{\xi} &=& \frac{1}{Q} \left [f'(H({\bf p},{\bf q})){\bf p}^T{\bf M}^{-1}{\bf p}-gk_BT \right ] \; . \label{gnh4} 
\end{eqnarray}
We can arrive at the 
natural expression for the momenta by performing the time
transformation
$d t/d\hat{\tau}=1/f'(H({\bf q},{\bf p}))$:
\begin{eqnarray}
d{{\bf q}}/d\hat{\tau} &=& {\bf M}^{-1}{\bf p} \label{gnh1a} \\
d{\bf p}/d\hat{\tau} &=&   -\nabla V({\bf q}) - \frac{1}{f'(H({\bf p},{\bf q}))}
\xi \bf p \label{gnh2a}\\
d{\eta}/d\hat{\tau} &=& \frac{1}{f'(H({\bf p},{\bf q}))}{\xi} \\
d{\xi}/d\hat{\tau} &=& \frac{1}{Q}\left [
{\bf p}^T{\bf M}^{-1}{\bf p}-\frac{1}{f'(H({\bf p},{\bf q}))}gk_BT \right ]
\;.  \label{gnh4a}
\end{eqnarray}
These equations have a  suggestive form.  The influence of the
modified distribution is manifested solely in the thermostat
variables.  Deviation of $f'(H)$ from unity can be viewed as a
time-dependent scaling of simulation temperature $T$ along with an inverse scaling
of the thermostat mass $Q$. 

It is possible to rewrite these equations yet again to achieve
separation of the coordinates and momenta.
Just as the quantity in (\ref{E0}) is constant
along solutions of the NH equations,
the GDD equations (\ref{gnh1a})--(\ref{gnh4a}) conserve
the related quantity
\begin{equation}
E^f_0 = f(H({\bf q},{\bf p}))+ \frac{Q\xi^2}{2}+gk_BT\eta. \label{gnhE0}
\end{equation}
Making use of our assumption that $F$ and (and also $f$) is monotonic 
and hence one-to-one,
we can solve (\ref{gnhE0}):
$$
H({\bf q},{\bf p})=f^{-1}\left(E^f_0- \frac{Q\xi^2}{2}-gk_BT\eta  \right)
$$
and define a new function of $\eta$ and $\xi$
\begin{equation}\phi(\eta,\xi)=1/f'\left( f^{-1}\left(E^f_0-
    \frac{Q\xi^2}{2}-gk_BT\eta \right) \right) \label{phidef}
\end{equation}
so that the GDD equations become
\begin{eqnarray}
d{{\bf q}}/d\hat{\tau} &=& {\bf M}^{-1}{\bf p} \label{gnh1b} \\
d{\bf p}/d\hat{\tau} &=&   -\nabla V({\bf q}) - 
\phi(\eta,\xi)\xi \bf p \label{gnh2b}\\
d{\eta}/d\hat{\tau} &=& \phi(\eta,\xi){\xi}  \label{gnh3b}\\
d{\xi}/d\hat{\tau} &=& \frac{1}{Q} \left [{\bf p}^T{\bf M}^{-1}{\bf p}-
\phi(\eta,\xi)gk_BT \right ].  \label{gnh4b}
\end{eqnarray}
Equations (\ref{gnh1b})--(\ref{gnh4b}) introduce coupling
between the thermostat variables, but leave the coordinates and momenta
separated, allowing for efficient discretization schemes. 
A similar trick can also be used to simplify the numerical calculations
in the symplectic Nos\'{e}-Poincar\'{e} method.  
We discuss numerical methods in the appendix.

\section{Example Application: A Variable Temperature Distribution} \label{tpd}
The methods of this paper are very general in the sense that dynamical
simulations can be made to sample any smooth,
invertible density function $F({\bf p},{\bf q})$.
In this section we propose a particular distribution function both for the purpose
of demonstrating the methods of this paper and to outline a potentially useful method
for accelerating sampling in systems with high barriers.

As mentioned earlier, one way to enhance the sampling of systems that are not
ergodic on the time scale of standard simulation due to high barriers is 
to carry out the simulations at high temperature.  The original distribution can
the be recovered by reweighting the trajectory to compensate for the change in the
distribution.  However, for most situations the low energy configurations that
have large weight in the original distribution are not sampled with sufficient
frequency at high temperatures to yield adequate statistics after reweighting. 

To address the low energy sampling problem with high temperatures, we
propose a generalized distribution that has the effect of raising
temperature only in high energy regions while leaving the low energy
dynamics unaffected.  To begin, we define the monotonic function 
$f_{\gamma}(s)$, designed to 
smoothly switch
between the identity function and a linear function of slope $\gamma$: 
\begin{eqnarray}
f_{\gamma}(s)= \left \{ \begin{array}{ll}
s&\mbox{   if   } s<s_0\\ 
as^3+bs^2+cs+d &\mbox{   if   } s_0\le s \le s_1\\ 
\delta+\gamma(s-s_1)&\mbox{   if   } s>s_1
  \end{array}   \right. \label{fofs}
\end{eqnarray}
where $s_0 \le \delta \le s_1$ control the size of the switching window and
the shape of the switching function, and the polynomial coefficients of
the switching function are given by
\begin{eqnarray*}
a&=& (2\delta - (1-\gamma)s_0 - (1+\gamma)s_1)/(s_0-s_1)^3   \\
b&=& (2(1-\gamma)s_0^2 + (2+\gamma)(s_0s_1 + s_1^2)- \nonumber \\
&& 3\delta(s_0+s_1))/(s_0-s_1)^3 \\
c&=& (\gamma s_0(s_0^2+s_0s_1-2s_1^2)- \nonumber \\
&& s_1(4s_0^2+s_0s_1-6\delta s_0 + s_1^2) )/(s_0-s_1)^3 \\
d&=& (s_0^2\delta(s_0-3s_1)+ \nonumber \\
&& s_1(2s_1+\gamma(s_1-s_0)))/(s_0-s_1)^3. 
\end{eqnarray*}
A graph of the function with $\gamma=0.2$, $s_0=3$, $s_1=4$, and
$\delta=(s_0+s_1)/2$ is shown in the left view of Figure~\ref{hdists}.

We now define a probability distribution as a function the Hamiltonian $H$:
\begin{equation}
F({\bf p},{\bf q}) = F_\gamma(H({\bf p},{\bf q}))=e^{-f_{\gamma}(H)/kT}. \label{vtemp}
\end{equation}
The same switching function $f_{\gamma}$ can be used to generate a separable distribution
with 
\begin{equation}
F({\bf p},{\bf q}) = e^{-{\bf p}^T {\bf M}^{-1} {\bf p}/2kT} 
e^{-f_{\gamma}(V({\bf q}))/kT} 
\end{equation}
in which the momentum distribution remains canonical. 
At $\gamma=1$ (and $\delta=s_1$) 
both the Hamiltonian and potential versions of the variable temperature
distribution reduce to the canonical distribution at temperature $T$. 
For $\gamma<1$,  the distributions give canonical 
sampling in low energy (total or potential, respectively)
regions ($E<E_0$ for some predetermined value $E_0$) of phase space at 
reference temperature $T$
while sampling high
energy regions ($E>E_1$) 
at the higher temperature 
$T(E)=T/\alpha(E)$ where
$\alpha(E)=\gamma(1-\frac{E_1}{E})+\frac{\delta}{E}$.
Note that
$\alpha(E)$ approaches $\gamma$ in the limit of large $E$.
This distribution modification has similarities with the Tsallis-based
distributions used by Andricioaei and Straub\cite{Andricioaei97}  and
Plastino and Anteneodo\cite{Plastino97} in that the effective temperature
is a monotonically increasing function of energy; however, in these cases
the temperature (and thus the dynamics) is altered at all energies whereas in our present case
the dynamics at low energies is unaltered.

\section{Numerical Experiments} \label{numexp}
To test our methods, we consider the double well potential
$$ V(x)=\epsilon(x^4-2x^2+1)$$
with minima at $x=\pm1$ and barrier height $\epsilon$.  

We have performed a number of GDD simulations using the 
variable temperature distribution (\ref{vtemp}) 
in both its full Hamiltonian and potential forms.
For all simulations, the reference 
temperature was $kT=\frac{\epsilon}{10}$.
For such low dimensional systems it is necessary to use Nos\'e-Hoover
chains to enhance the ergodicity of the dynamics --- see the Appendix
for discussion. In all simulations six thermostats were used.  The switching
window parameters in (\ref{fofs}) were taken as $s_0=3$, $s_1=4$ and
$\delta=3.5$ (except when $\gamma=1$, then $\delta=s_1$).  
We report experiments with the variable temperature
parameter $\gamma$=1.0, 0.9, 0.8, 0.7, 0.6, 0.5, 0.4, 0.3, 0.2.
As $\gamma$ decreases from unity, stability
considerations dictate smaller timesteps.
The timestep
was $h=0.001$ for the modified potential energy calculations.
For the full Hamiltonian approach,
we use timestep $h=0.0001$.

\begin{figure}
\rotatebox{90}{\includegraphics[scale=0.3]{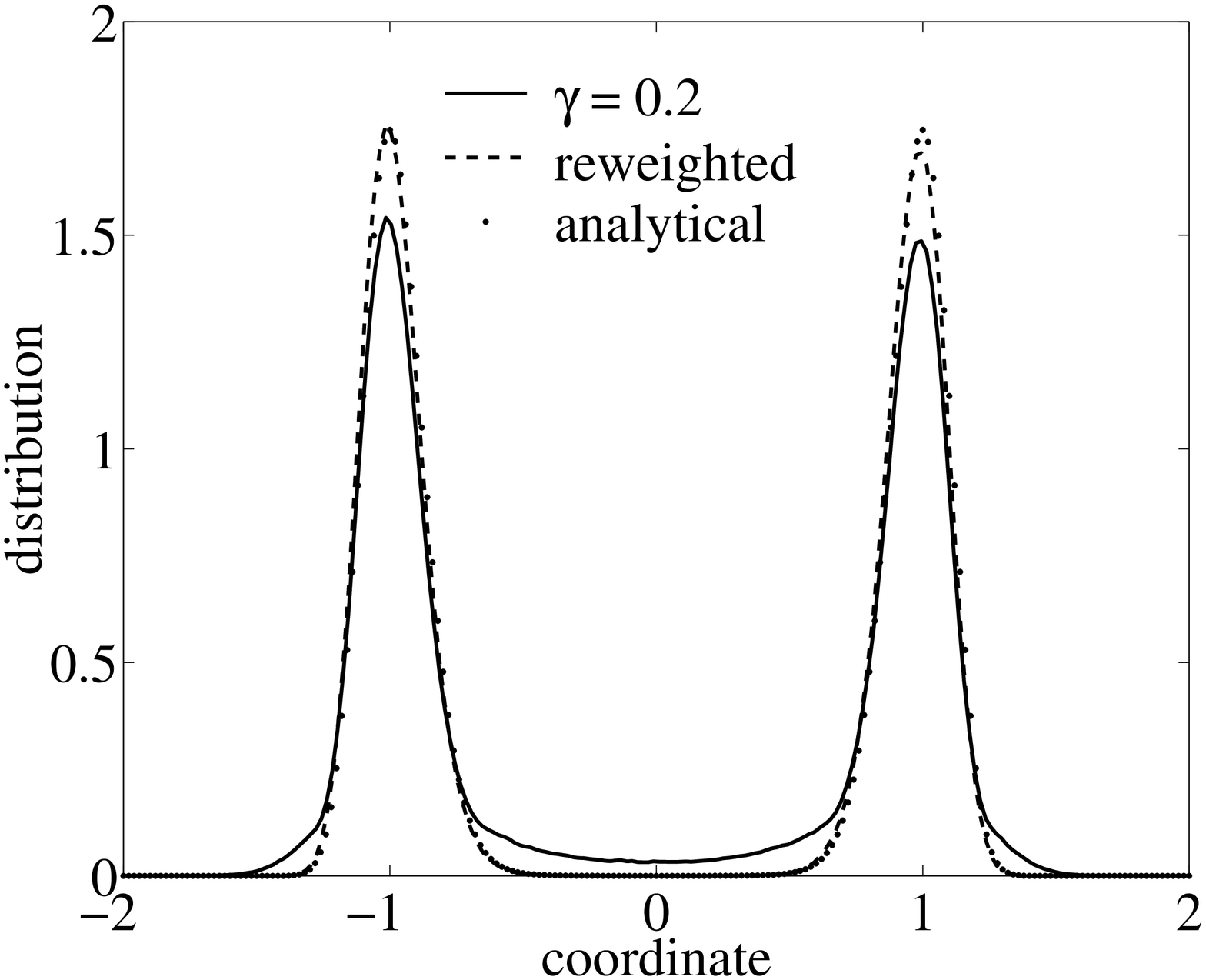}}
\rotatebox{90}{\includegraphics[scale=0.3]{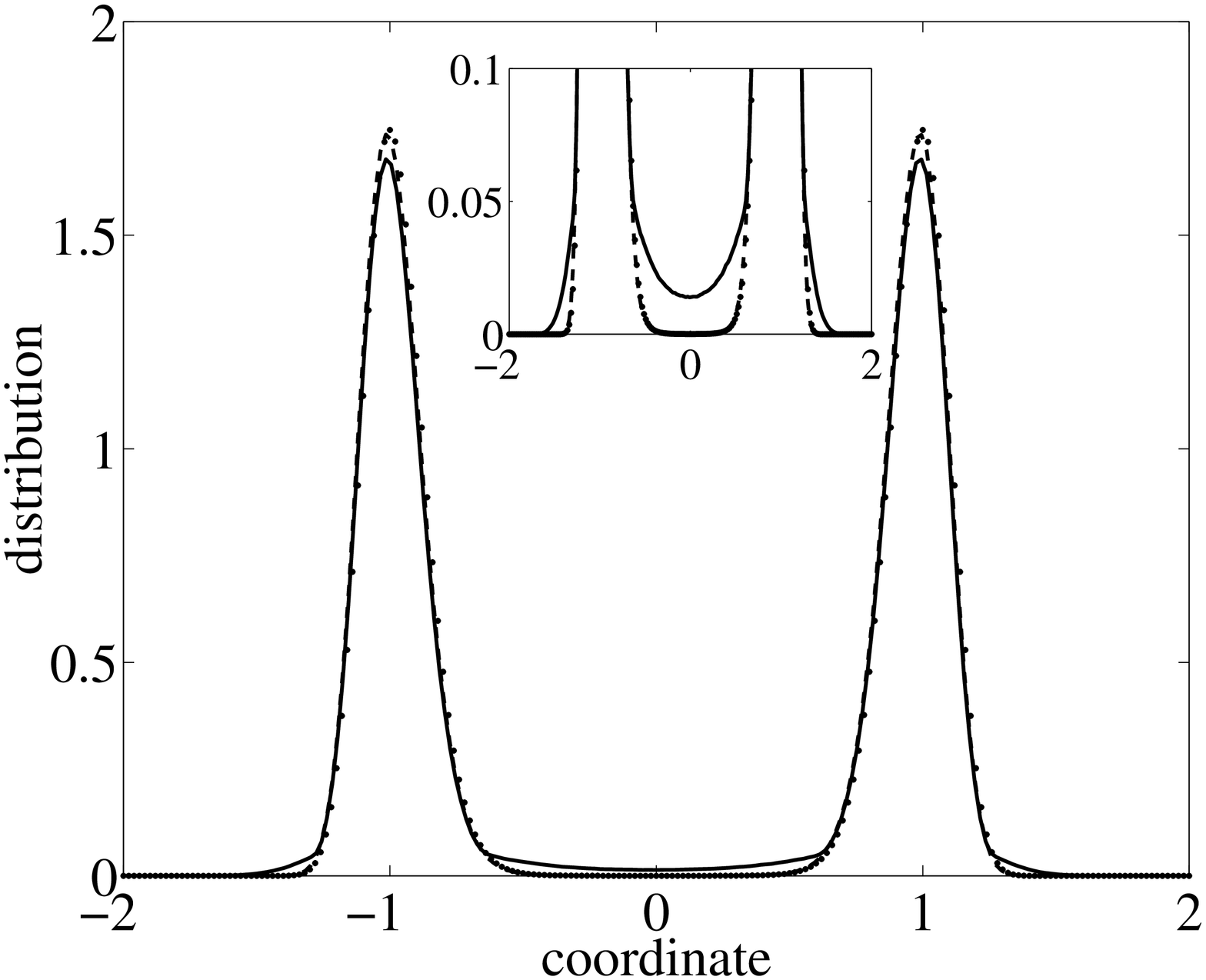}}
\caption{Top view: Full Hamiltonian simulation $\rho_\gamma$ with $\gamma=0.2$,
Bottom view:  Modified Potential simulation with $\gamma=0.2$. 
The solid curve gives the coordinate distribution of the computed trajectories,
the dashed line gives the reweighting of the computed sampling to the 
canonical distribution, and the dotted lines give the theoretical canonical distribution.
\label{xdists}}
\end{figure}

The left view of Figure~\ref{xdists} shows the distribution of coordinates for
the full Hamiltonian calculations at $\gamma=0.2$.
It can be seen by the
good agreement with the theoretical distribution 
that the
coordinates along the trajectory are sampled according to $F_\gamma$.
Also shown is the reweighted
distribution which recovers the canonical distribution.
Note that at this temperature ($kT=\epsilon/10$) standard NH chain
dynamics fails to sample effectively.  
It must be pointed out that because 
the GDD equations for the full Hamiltonian  case
were derived with a time transformation
$d t /d \hat{\tau} = \phi(\eta,\xi)$ in (\ref{phidef}), it is necessary to include
$\phi$ as a weighting function when computing averages using
trajectories produced by equations (\ref{gnh1b})--(\ref{gnh4b}).
The right view of Figure~\ref{xdists} shows 
the distribution of coordinates for
the modified potential calculations at $\gamma=0.2$.
As for the results for the full Hamiltonian, it can be seen that the canonical 
coordinate distribution is also recovered by reweighting.  Note that the unweighted coordinate 
distributions from the full Hamiltonian and modified potential
formulations are not identical.  In particular, the trajectory from the 
full Hamiltonian method spends slightly more time in high energy 
configurations.

\begin{figure}
\rotatebox{90}{\includegraphics[scale=0.30]{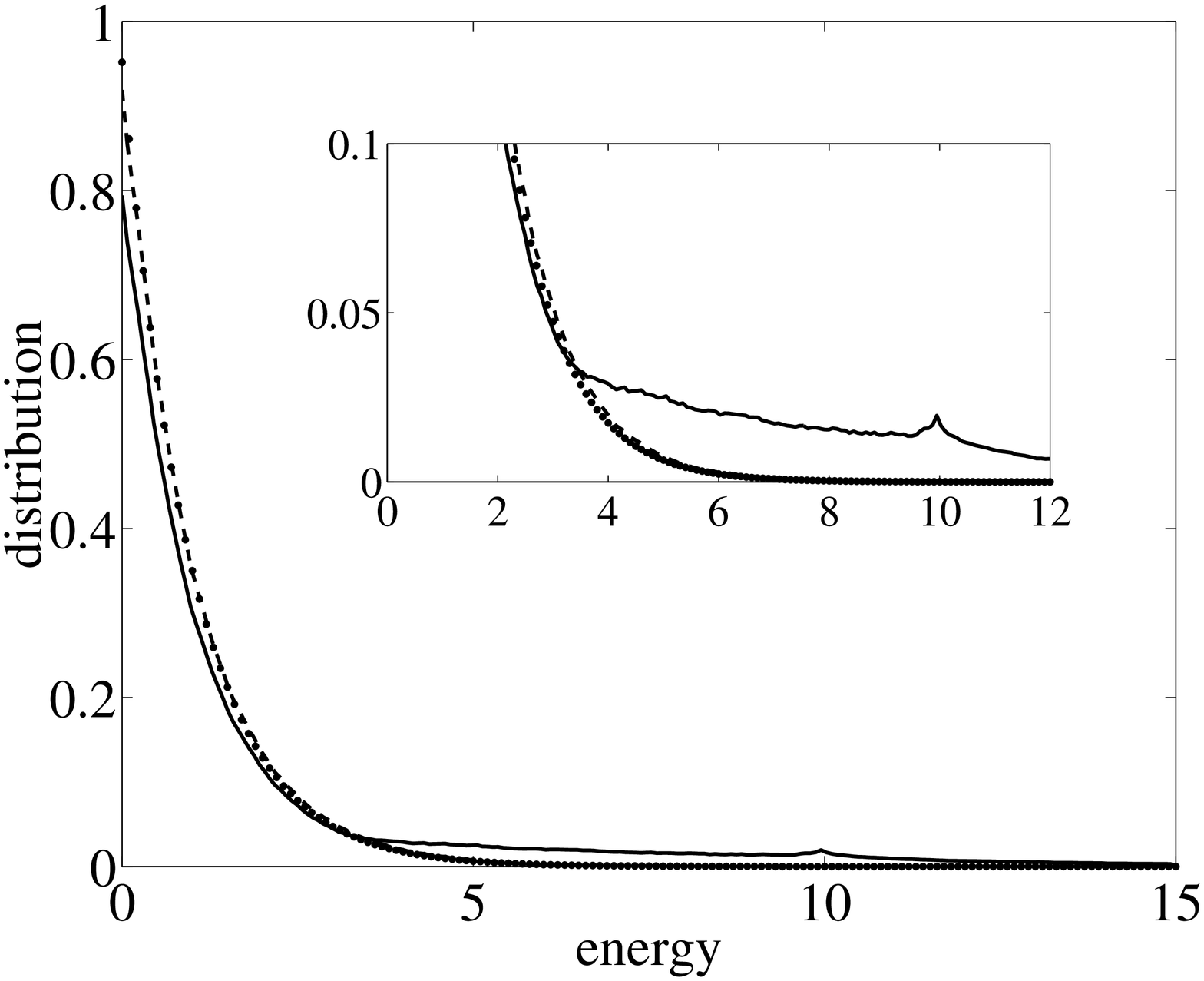}}
\rotatebox{90}{\includegraphics[scale=0.30]{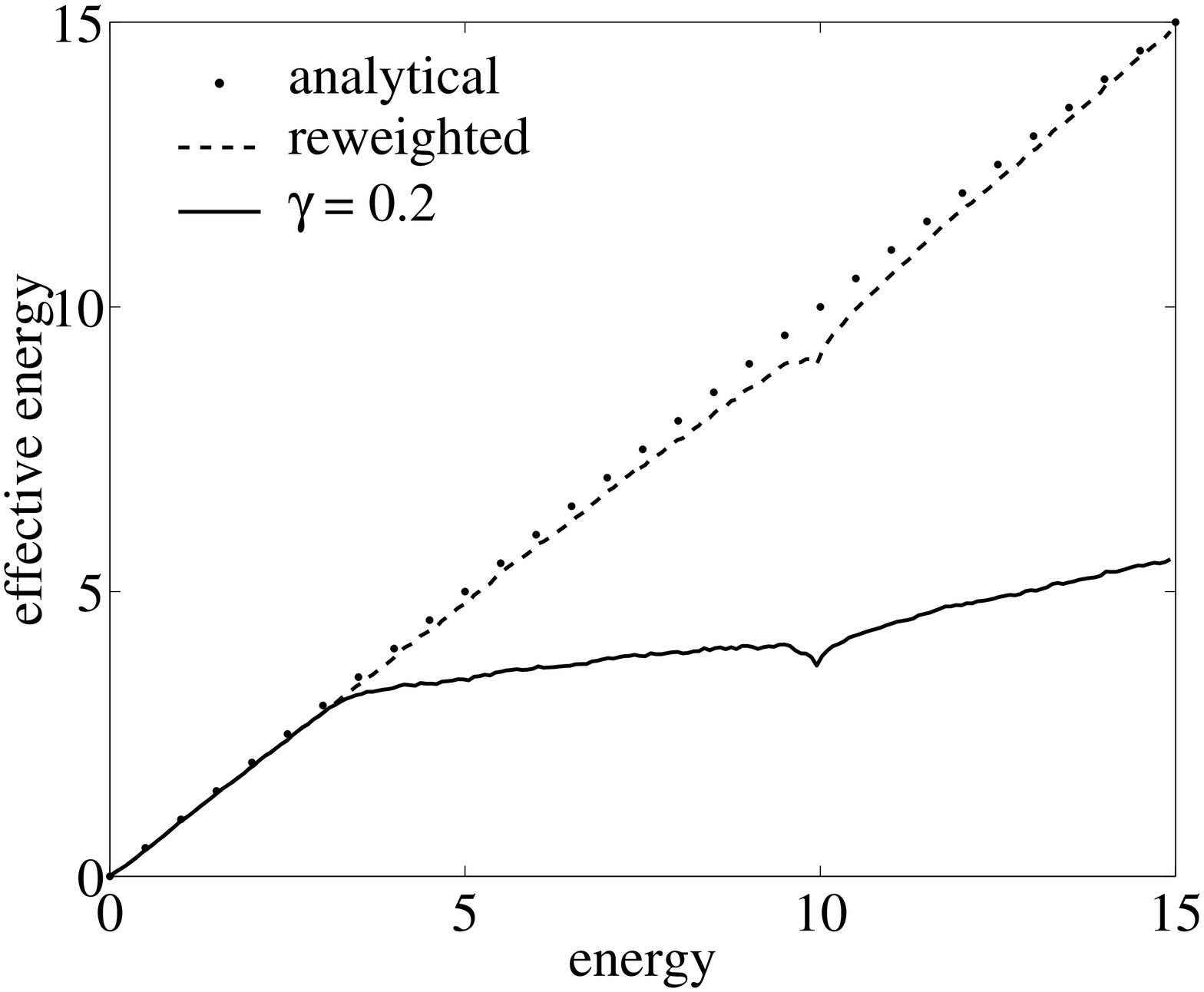}}
\caption{Top View:  Total energy density for the Full Hamiltonian
distribution;  Bottom View:  The function $f(H)$ reconstructed from 
trajectory energy sampling. 
The solid curve gives the total energy distribution of the computed trajectories,
the dashed line gives the reweighting of the computed sampling to the 
canonical distribution, and the dotted lines give the theoretical canonical distribution.
\label{hdists}}
\end{figure}

In Figure~\ref{hdists} we show the distribution of total energy 
along the computed full Hamiltonian trajectory 
for $F_\gamma$ with $\gamma=0.2$,
which can be seen to closely approximate the theoretical energy
distribution for $F_\gamma$.  Also shown  is the
function $f(H)$ from equation~(\ref{fofs}), along with
the computed values of $f$
calculated as the
natural logarithm of 
the energy distribution from the trajectory.

\begin{figure}
\centerline{
\rotatebox{90}{\includegraphics[scale=0.35]{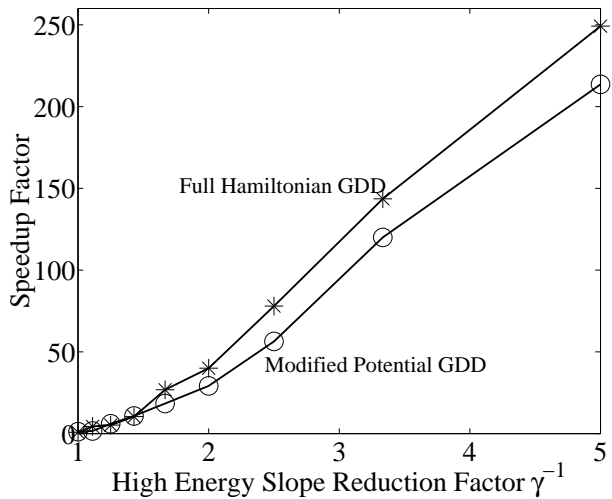}}}
\caption{Sampling speedup measured as increased frequency of barrier crossings
for the full Hamiltonian (stars) and modified potential (circles)  
methods using the variable temperature 
distribution.  
The high energy slope reduction factor is $\gamma^{-1}$.  The speedup factor
is the average waiting time between barrier crossings for each value of $\gamma$,
normalized by the average waiting time at $\gamma=1$.
\label{waiting}}
\end{figure}

In Figure~\ref{waiting} we illustrate the success of the variable
temperature density $F_\gamma$ in hastening barrier crossings
for the double well system.  The figure shows waiting time plotted versus
the temperature boost factor $1/\gamma$.  It can be seen that both
the full Hamiltonian and modified potential
approaches yield dramatic reductions in waiting time between barrier
crossings.

\section{Conclusion} \label{conclusion}
In this paper we have presented a general dynamical formalism, 
which we call Generalized Distribution Dynamics (GDD), that
generates trajectories according to any of a broad class of probability 
distribution functions.  In addition we show that the GDD scheme, which is based on the 
Nos\'e thermostat\cite{Nose84a}, can be easily implemented numerically  for two classes
of distribution functions: distributions that are functions of the full Hamiltonian
and those that separate into a product of momentum and position distributions. 
In these two cases, the GDD scheme is equivalent to the dynamics of a system
with a modified full Hamiltonian or effective potential energy surface, respectively.
To implement GDD for these two classes, we outline specific numerical 
methods for both the Nos\'e-Hoover\cite{Hoover85} and 
Nos\'e-Poincar\'e\cite{Bond99} real-time formulations of Nos\'e dynamics.
As an example, we have introduced a specific form of a probability density function,
the Variable Temperature Distribution, which has application in accelerating configurational
sampling in systems with high energy barriers.
To illustrate the numerical scheme and evaluate the method we performed 
numerical experiments using a one-dimensional bistable
oscillator and demonstrate that the Variable Temperature Distribution is
very effective for accelerated sampling of coordinates
when used in the effective potential energy setting.  
We are currently applying this work to enhance the dynamical sampling of 
model polypeptides.

\section{Acknowledgements}
This work was undertaken during visits of EJB and BBL to the Centre for 
Mathematical Modelling at the University of Leicester, UK.
Acknowledgment is made by EJB to the donors of The Petroleum Research Fund, administered by 
the ACS, for partial support of this research.  BBL acknowledges the support
of the National Science Foundation
under grant CHE-9970903. BJL acknowledges the UK Engineering and Physical Sciences Research Council grant GR/R03259/01.

\renewcommand{\theequation}{A-\arabic{equation}} 
\setcounter{equation}{0}  
\renewcommand{\thesubsection}{A-\arabic{subsection}} 
\setcounter{subsection}{0}  
\renewcommand{\thesubsubsection}{A \arabic{subsubsection}} 
\setcounter{subsubsection}{0}  

\section*{Appendix}  

\subsection{Numerical Methods for Nos\'e-Hoover GDD for distributions that are functions
of the Hamiltonian}
The NH equations (\ref{nh1})--(\ref{nh4}) can be discretized in a number of ways which
generalize the basic Verlet approach.  
One such method is \cite{Martyna96}:
\begin{eqnarray}
{\bf p}_{n+\frac{1}{2}}&=&{\bf p}_{n}-\frac{\Delta t}{2}\left(\nabla V({\bf q}_n) + {\xi_{n+\frac{1}{2}}}{\bf  p}_{n+\frac{1}{2}}\right)\label{s1} \\
\xi_{n+\frac{1}{2}}&=&\xi_{n}+\frac{\Delta t}{2Q}\left({\bf p}_{n+\frac{1}{2}}^T{\bf M}^{-1}{\bf p}_{n+\frac{1}{2}}-\right . \nonumber \\
&& \left . gk_BT\right)\label{s2} \\
\eta_{n+1}&=&\eta_n+\Delta t \,{\xi_{n+\frac{1}{2}}}\label{s3}\\
{\bf q}_{n+1}&=&{\bf q}_{n}+\Delta t \, {\bf M}^{-1}{\bf p}_{n+\frac{1}{2}}
\label{s4} \\
\xi_{n+1}&=&\xi_{n+\frac{1}{2}}+\frac{\Delta t}{2Q}\left({\bf p}_{n+\frac{1}{2}}^T{\bf M}^{-1}{\bf p}_{n+\frac{1}{2}}-\right . \nonumber \\
&& \left . gk_BT\right)\label{s5} \\
{\bf p}_{n+1}&=&{\bf p}_{n+\frac{1}{2}}-\frac{\Delta t}{2}\left(\nabla V({\bf q}_{n+1}) \right .+ \nonumber \\
&& \left .{\xi_{n+\frac{1}{2}}}{\bf  p}_{n+\frac{1}{2}}\right).\label{s6}
\end{eqnarray}
Equations (\ref{s1}) and (\ref{s2}) can be solved by computing the
vector
$$\bar{\bf p}={\bf p}_{n}-\frac{\Delta t}{2}\nabla V({\bf q}_n) $$
and rewriting (\ref{s2})
$$\xi_{n+\frac{1}{2}}=\xi_{n}+\frac{\Delta t}{2Q}\left(\frac{\bar{\bf p}^T{\bf M}^{-1}\bar{\bf p}}{\left(1+\frac{\Delta t}{2}{\xi_{n+\frac{1}{2}}}\right)^2}-gk_BT \right).$$
This scalar equation for  $\xi_{n+\frac{1}{2}}$ can now be solved analytically or with
an iterative solver.  Using the computed $\xi_{n+\frac{1}{2}}$, we can compute
$${\bf p}_{n+\frac{1}{2}} = \frac{\bar{\bf p}}{1+\frac{\Delta t}{2}{\xi_{n+\frac{1}{2}}}}.$$

The discretization scheme (\ref{s1})--(\ref{s6}) can be generalized
to the  GDD equations (\ref{gnh1b})--(\ref{gnh4b}):
\begin{eqnarray}
{\bf p}_{n+\frac{1}{2}}&=&{\bf p}_{n}-\frac{\Delta t}{2}\left(\nabla V({\bf
    q}_n)\right . + \nonumber \\
&& \left .\phi(\eta_n,\xi_{n+\frac{1}{2}}){\xi_{n+\frac{1}{2}}}{\bf  p}_{n+\frac{1}{2}}\right)\label{gs1} \\
\xi_{n+\frac{1}{2}}&=&\xi_{n}+\frac{\Delta t}{2Q}\left({\bf p}_{n+\frac{1}{2}}^T{\bf M}^{-1}{\bf p}_{n+\frac{1}{2}}-\right . \nonumber\\
&& \left .\phi(\eta_n,\xi_{n+\frac{1}{2}})gk_BT\right)\label{gs2} \\
\eta_{n+\frac{1}{2}}&=&\eta_n + \frac{\Delta t}{2}\phi(\eta_n,\xi_{n+\frac{1}{2}}){\xi_{n+\frac{1}{2}}}\label{gs3}\\
{\bf q}_{n+1}&=&{\bf q}_{n}+\Delta t {\bf M}^{-1}{\bf p}_{n+\frac{1}{2}}
\label{gs4} \\
\eta_{n+1}&=&\eta_{n+\frac{1}{2}}+ \frac{\Delta t}{2}\phi(\eta_{n+1},\xi_{n+\frac{1}{2}}){\xi_{n+\frac{1}{2}}}\label{gs5}\\
\xi_{n+1}&=&\xi_{n+\frac{1}{2}}+\frac{\Delta t}{2Q}\left({\bf p}_{n+\frac{1}{2}}^T{\bf M}^{-1}{\bf p}_{n+\frac{1}{2}}- \right .\nonumber \\
&&\left . \phi(\eta_{n+1},\xi_{n+\frac{1}{2}})gk_BT\right)\label{gs6} \\
{\bf p}_{n+1}&=&{\bf p}_{n+\frac{1}{2}}-\frac{\Delta t}{2}\left(\nabla V({\bf
    q}_{n+1}) \right . + \nonumber \\
&& \left .\phi(\eta_{n+1},\xi_{n+\frac{1}{2}}){\xi_{n+\frac{1}{2}}}{\bf  p}_{n+\frac{1}{2}}\right).\label{gs7}
\end{eqnarray}
Two of the formulae are implicit.  The procedure for
$\xi_{n+\frac{1}{2}}$ is the same as for (\ref{gs2}), while (\ref{gs5}) may
require
the use of an iterative method, depending on the nature of the
function
$\phi$.  In the numerical results presented here, we use Newton-Raphson
iteration with tolerance of $10^{-12}$ in double precision.

The GDD scaling function $\phi(\eta,\xi)$ introduced in equation (\ref{phidef})
serves two interesting purposes.  The resulting dynamical formalism
retains the form of the NH equations, with coordinates and momenta coupled
to a generalized thermostat (subject to a time transformation) via
the thermostat variables $\eta$ and $\xi$.
From a practical point of view the introduction of $\phi$ allows for
discretization schemes which are explicit in the coordinates and momenta,
which is important for overall efficiency.  Implementation of 
a timestepping scheme such as (\ref{gs1})--(\ref{gs7}) requires repeated
evaluation of the function $\phi$, requiring the inversion of the function
$f$ in (\ref{Heff}) which defines the effective Hamiltonian of the generalized
density.  In general an analytic expression will not be available
for $f^{-1}$. For the work described here, we have implemented the 
GDD scaling function using an algorithm
which relies on the Newton-Raphson method for finding a zero of a scalar
equation.  For the variable temperature distribution based on (\ref{fofs}),
we evaluate $\phi(\eta,\xi)=1/f'(f^{-1}(H))$ by first evaluating 
$ \Delta E = E^f_0-\frac{Q \xi^2}{2}-gk_BT\eta$ from equation 
(\ref{phidef}), then evaluating
$$f^{-1}(\Delta E)=\left \{ \begin{array}{ll}
\Delta E &\mbox{   if   } \Delta E<H_0\\ 
\mbox{root of} as^3+bs^2+\nonumber \\
 cs+d - \Delta E &\mbox{   if   } H_0\le \Delta E \le H_1\\ 
\frac{1}{\gamma}(\Delta E-\delta)+H_1&\mbox{   if   } \Delta E>H_1
  \end{array}   \right., \label{phiproc}$$
and 
$$f'(f^{-1})
=\left \{ \begin{array}{ll}
1 &\mbox{   if   } \Delta f^{-1}<H_0\\ 
3as^2+2bs+c &\mbox{   if   } H_0\le f^{-1} \le H_1\\ 
\gamma &\mbox{   if   } f^{-1}>H_1
  \end{array}   \right.. \label{phiproc2}
$$
With a good initial approximation (such as $\Delta E$ 
or the average $\frac{1}{2}(\Delta E + \frac{1}{\gamma}(\Delta E-\delta)+H_1)$)
the Newton-Raphson method above
converges to the desired root with
two or three iterations in our experience, subject to a convergence tolerance
of $10^{-12}$ in double precision calculations.

\subsection{A symplectic numerical method for Nos\'{e}-Poincar\'{e} GDD for distributions
that are functions of the Hamiltonian }
Starting from the Nos\'{e} extended
Hamiltonian applied to the effective Hamiltonian, 
\begin{equation}
H_{Nos \acute{e}}^f = -f(H(q,\tilde{p}/s)) + \frac{\pi^2}{2Q} + gk_BT  \ln s
\end{equation}
Assuming $f$ is one to one, we can invert $f$ to obtain, along the energy
surface $H_{\rm Nose}^f = E$, the new Hamiltonian
\begin{equation}
H(q,\tilde{p}/s) -f^{-1}( E_0^f - \frac{\pi^2}{2Q} - g k T \ln s) = 0.
\end{equation}
The zero energy dynamics in this Hamiltonian correspond to 
$H_{\rm Nose} = E_0^f$ dynamics.  We next introduce a 
time-transformation of Poincar\'{e} type, $H\rightarrow sH$
resulting in 
\begin{eqnarray}
H_{NP}^f &=& sH(q,\tilde{p}/s) - sf^{-1}( E_0^f - \nonumber \\
&& \frac{\pi^2}{2Q} - g k T \ln s) = 0.
\end{eqnarray}
It is natural to use a splitting method here, breaking the Hamiltonian into two
parts according to the obvious additive decomposition and solving each term successively using an appropriate symplectic numerical method.  Note that the splitting suggested here is different than that used recently by Nos\'{e} \cite{Nose00} in his variation of the Nos\'{e}-Poincar\'{e} method, but the basic technique is similar.  Integration of the term 
\begin{equation}
H_1 = sH(q,\tilde{p}/s)
\end{equation}
can be easily performed using the standard (and symplectic) Verlet method; note that during this fraction of the propagation timestep, $s$ will be constant.
Formally, the integration of 
\begin{equation}
H_2 = -sf^{-1}( E_0^f - \frac{\pi^2}{2Q} - g k T \ln s) 
\end{equation}
can be done analytically.  However, this is relatively painful.  A simpler
approach is to use an implicit method, such as the implicit midpoint method. For a general Hamiltonian $H(q,p)$, the midpoint method advances from step to
step by solving
\begin{eqnarray}
{\bf q}_{n+1} & = {\bf q}_n + \Delta t \nabla_p H({\bf q}_{n+\frac{1}{2}},{\bf p}_{n+\frac{1}{2}}) \\ 
{\bf p}_{n+1} & = {\bf p}_n + \Delta t \nabla_q H({\bf q}_{n+\frac{1}{2}},{\bf p}_{n+\frac{1}{2}})
\end{eqnarray}
where
${\bf q}_{n+1/2} \equiv ({\bf q}_n + {\bf q}_{n+1})/2$ and ${\bf p}_{n+1/2}$
is defined similarly.
In the present case, this means solving a nonlinear system in ${\bf R}^2$ at
each timestep.

\subsection{Nos\'e-Hoover chains}
For systems that are small or contain stiff oscillatory components,
lack of ergodicity may render the Nos\'e scheme ineffective.
Chains of Nos\'e-Hoover thermostats have been shown\cite{Martyna92} to allow canonical sampling
in these cases.  
For a chain
of $m+1$ thermostat variables the equations of motion are
\begin{eqnarray}
d{{\bf q}}/d\tau &=& {\bf M}^{-1}{\bf p} \label{nhc1} \\
d{\bf p}/d\tau &=&   -\nabla V({\bf q}) - {\xi_0}{\bf p} \label{nhc2}\\
d{\eta_0}/d\tau &=& {\xi_0}  \label{nhc3}\\
d{\xi_0}/d\tau &=& \frac{1}{Q_0}\left[{\bf p}^T{\bf M}^{-1}{\bf p}-gk_BT\right]-{\xi_1}\xi_0  \label{nhc4}\\
d{\eta_1}/d\tau &=& {\xi_1}  \label{nhc5}\\
d{\xi_1}/d\tau &=& \frac{1}{Q_1}\left[{Q_0\xi_0^2}-k_BT\right]-{\xi_2}\xi_1  \label{nhc6}\\
& \vdots&  \nonumber \\
d{\eta_{m-1}}/d\tau &=& {\xi_{m-1}}  \label{nhc7}\\
d{\xi_{m-1}}/d\tau &=& \frac{1}{Q_{m-1}}\left[Q_{m-2}{\xi_{m-2}^2}-k_BT\right]-{\xi_m}\xi_{m-1}  \label{nhc8}\\
d{\eta_m}/d\tau &=& {\xi_m}  \label{nhc9}\\
d{\xi_m}/d\tau &=& \frac{1}{Q_m}\left[Q_{m-1}{\xi_{m-1}^2}-k_BT\right].  \label{nhc10}
\end{eqnarray}
Along solutions of the Nos\'e-Hoover Chain (NHC) equations the conserved quantity is
\begin{eqnarray}
E_{NHC} &=& \frac{{\bf p}^T{\bf M}^{-1}{\bf p}}{2} + V({\bf
  q})+ \sum_{i=0}^m\frac{Q_i\xi_i^2}{2}+\nonumber \\
&& gk_BT \eta_0 +
\sum_{i=1}^mk_BT \eta_i . \label{EC0}
\end{eqnarray}
Discretization schemes for NH chains are discussed in References \cite{Jang97} and \cite{Martyna92}.  
Extension of Nos\'{e}-Poincar\'{e} to incorporate chains is somewhat delicate; this is work in progress 
by two of the authors.

\end{document}